\documentclass[fleqn,twoside]{article}
\usepackage{espcrc2}
\usepackage{graphicx}

\newcommand{\beq}{\begin{equation}}
\newcommand{\eeq}{\end{equation}}
\newcommand{\bea}{\begin{eqnarray}}
\newcommand{\eea}{\end{eqnarray}}
\newcommand{\bweq}{\begin{widetext}\begin{equation}}
\newcommand{\eweq}{\end{equation}\end{widetext}}
\newcommand{\bwt}{\begin{widetext}}
\newcommand{\ewt}{\end{widetext}}

\newcommand{\scs}[1]{\mbox{\scriptsize{#1}}}

\newcommand{\m}{\mbox{\ m}}

\newcommand{\km}{\mbox{\ km}}

\newcommand{\MeV}{\mbox{\ MeV}}
\newcommand{\GeV}{\mbox{\ GeV}}
\newcommand{\TeV}{\mbox{\ TeV}}

\newcommand{\EeV}{\mbox{\ EeV}}

\newcommand{\fmc}{\mbox{\ fm/c}}
\newcommand{\gcmIS}{\mbox{\ g$\cdot$cm$^{-2}$}}

\begin{document}

\title{On Sensitivity of Cherenkov Radiation to the Dynamics of High
  Energy Cosmic Ray Interactions}
\author{J.\ \v R\' \i dk\' y 
 \address{Institute of Physics of the Academy of Sciences of the
          Czech Republic, \\
          Na Slovance 2, 182 21 Prague 8, Czech Republic}
  \thanks{ridky@fzu.cz}
          and
        D.\ Nosek
 \address{Institute of Particle and Nuclear Physics, Charles
         University, \\
         V Hole\v sovi\v ck\'ach 2, 180 00 Prague 8, Czech Republic}
  \thanks{nosek@ipnp.troja.mff.cuni.cz}
}

\begin{abstract}
A simulation study of the effects resulting from creation of
quark--gluon plasma in high energy collisions of cosmic iron nuclei
with the air has been carried out.
The Cherenkov light emission has been found to reveal some features 
of the dynamics of high energy hadronic interactions.
\end{abstract}

\maketitle

\section{Introduction}
\label{IN}

One possible way to investigate rare ultra--energetic collisions of cosmic 
ray primaries with the air is to investigate the Cherenkov 
light emitted in atmospheric showers.
Charged daughter particles produced along a hadronic cascade have
mostly very high velocities and therefore emit in the air Cherenkov 
photons which in principle carry an information about longitudinal shower
development including an information on the characteristics of the 
primary collision, e.g. particle multiplicity. 
As optical photons suffer relatively smaller absorption in the atmosphere, 
their density on the ground is higher than that of charged 
particles and their lateral distribution is broader.
The Cherenkov light highlights differences in the mechanism
of hadronic collisions that are expected to occur in the first 
stage of the shower development and that can propagate 
further during development of the cascade.

\section{Primary Interaction}
\label{FI}

Since available experimental data do not contradict to the fact that 
the abundance of $^{56}$Fe nuclei in the primary flux may be
above $10\%$~\cite{Swo1}, a non--negligible portion of iron--nitrogen 
or iron--oxygen nucleus--nucleus collisions can be observed in upper 
atmosphere.
The number of nucleons participating in such collisions may well 
grow up to a few tens and a fireball of hot and dense elementary
matter of quarks and gluons (QGP) can be created.
Its thermal equilibrium is achieved in rapid statistical energy 
distribution among partons.
As a result an almost ideally thermalized system of nucleons, pions
and kaons (heavier baryons and mesons are neglected in our approach) 
is produced in the primary collision.

\begin{figure}[h!]
\vspace{-1.7cm}
\centerline{
\includegraphics[width=7.5cm]{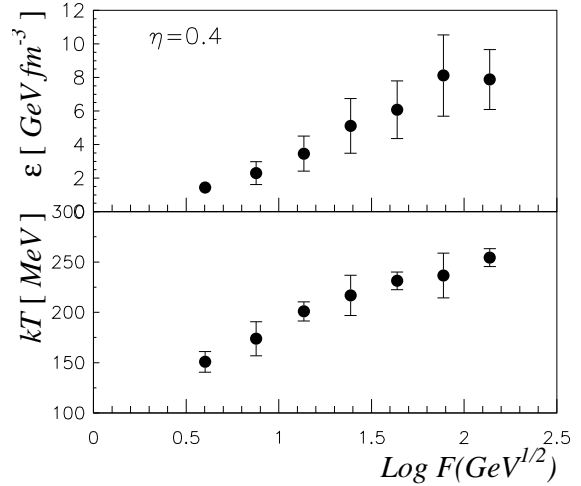}}
\vspace{-1.2cm}
\caption{Estimated energy densities and transition temperatures 
  achieved at the first Fe--N interaction are depicted as functions 
  of the Fermi energy variable (see text).}
\label{F01}
\vspace{-1.0cm}
\end{figure}

The energy density of the fireball can be estimated from the 
Bj\o rken formula~\cite{Bjo1}, 
$\epsilon = \frac{1}{\pi R^{2} \tau} \frac{\Delta E}{\Delta y}$,
where $R$ is the radius of the smaller nucleus, the formation time 
was chosen as $\tau \approx 1 \fmc$ and $\Delta E$ is c.m.s. energy 
of produced particles in the unit rapidity interval in the vicinity of 
its c.m.s. value.
Energy densities and corresponding upper bounds of the transition 
temperatures achieved in our simulations of the primary interaction 
of the cosmic iron nucleus with the air nitrogen are shown in
Fig.\ref{F01} as functions of the Fermi energy variable 
$F \approx (s_{NN})^{\frac{1}{4}}$, where $\sqrt{s_{NN}}$ is an 
invariant mass of a pair of interacting nucleons~\cite{Gaz1}.

\begin{figure}[h!]
\vspace{-2.8cm}
\centerline{
\includegraphics[width=7.5cm]{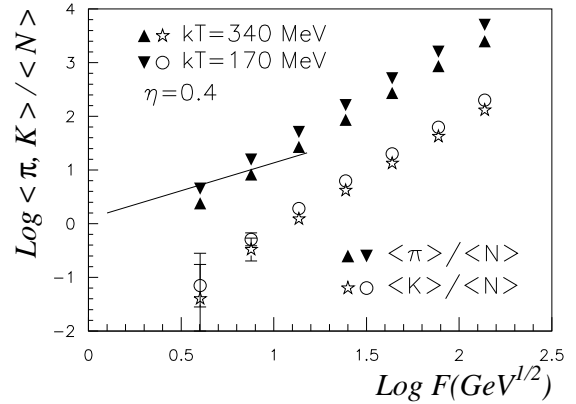}}
\vspace{-1.0cm}
\caption{The average yields of pions and kaons per average number 
  of participating nucleons as functions of the Fermi energy variable. 
  SPS and RHIC experimental results for pions are indicated by the 
  straight line.}
\label{F02}
\vspace{-1.2cm}
\end{figure}

When the number of wounded nucleons $N_{\scs{Int}}$ in the first
collision is greater than a certain value 
(typically $N_{\scs{Int}} \ge 20$), we assume that 
several nucleons melt down into the QGP phase and
a certain part of the total mass of the fireball 
$E_{\scs{hot}} \approx \eta \sqrt{s_{\scs{F}}}$ is released. 
The model parameter $\eta$ ($\eta = 0.1-0.4$) is assumed to 
be independent of the collision energy and the system size.
A portion of released energy is channeled to nucleons with 
momenta generated according to the relativistic Boltzmann 
distribution isotropically with respect to the fireball 
center of mass system.
The remaining energy is then used to create a number of $\pi$ and $K$ 
mesons with isotropic and relativistic Boltzmann distributed 
momenta in fireball system. 
Thermalized hadronic species are produced with probabilities 
$N : \pi : K \approx 10 : 85 : 5$ and with approximately equal 
abundances of different charges.
Chemical freeze--out is controlled by a constant temperature 
($kT = 170, 340 \MeV$) and zero chemical potentials. 
Multiplicities of light mesons produced in the initial interaction, 
as calculated in our model, are shown in Fig.\ref{F02}.

\section{Shower Simulations}
\label{SS}

To achieve the quantitative results, the CORSIKA simulation
code~\cite{Hec1} has been adapted to calculate the contributions 
of the Cherenkov photon intensity stemming from the quark--gluon
equilibration in primary collisions of the cosmic iron nuclei 
with the air.
The QGSJET01 model of high energy interactions has been 
employed; the GEISHA procedure has been used to treat hadronic 
collisions at energies $E < 80 \GeV$.
It is worth to note that the uncertainty related to the modelling 
of standard hadronic interactions turns out to be of minor influence,
as the present investigation of QGP effects relies upon comparative 
characteristics of the detectable photons.

We analyze the contributions of Cherenkov light to the lateral
ground level distribution functions of iron--induced showers for 
a given primary energy ($E_{\scs{Fe}} = 100 \TeV - 10 \EeV$) 
averaged over 100 showers.
The primary zenith angle was fixed at zero degrees; the altitude of
the initial interaction was left free.
The kinetic energy cutoffs for hadrons and muons have been set 
to $0.3 \GeV$; for electrons/positrons and photons we use cutoffs 
of $20 \MeV$ and $2 \MeV$, respectively.
In all calculations a thinning level $10^{-7}$ has been adopted.
The Cherenkov light has been collected by a row of 100 ideal detectors 
with detection area $3\times 3 \m^2$ and spacing $10 \m$ 
located in north--south direction at a height of $1 \km$ above the sea
level ($813 \gcmIS$).
In all simulations the standard U.S. atmosphere has been used;
propagation of the Cherenkov light has been corrected for attenuation
in the air.

\section{Effects of QGP formation}
\label{QGP}

In our simple model, more secondary particles are created in 
the first interaction when QGP is formed (QGP shower)
compared to the ordinary Fe--N collisions (Fe shower). 
Air showers following QGP production have their maxima higher in 
the atmosphere and produce noticeably more Cherenkov light at higher 
altitudes; this effect grows with increasing incident energy.
\begin{figure}[h!]
\vspace{-1.7cm}
\centerline{
\includegraphics[width=7.5cm]{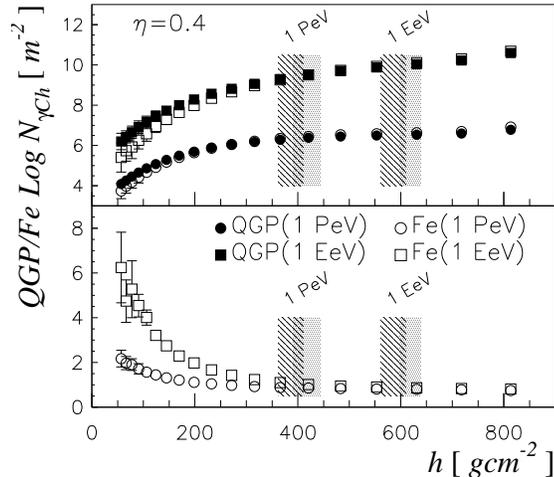}}
\vspace{-1.3cm}
\caption{Average ground level densities and relative QGP/Fe numbers 
of Cherenkov photons are depicted as functions of the slant 
depth of their production.
Hatched (dotted) boxes correspond to maxima of the QGP (Fe) showers.} 
\label{F03}
\vspace{-1.0cm}
\end{figure}
\begin{figure}[h!]
\vspace{-0.7cm}
\centerline{
\includegraphics[width=7.5cm]{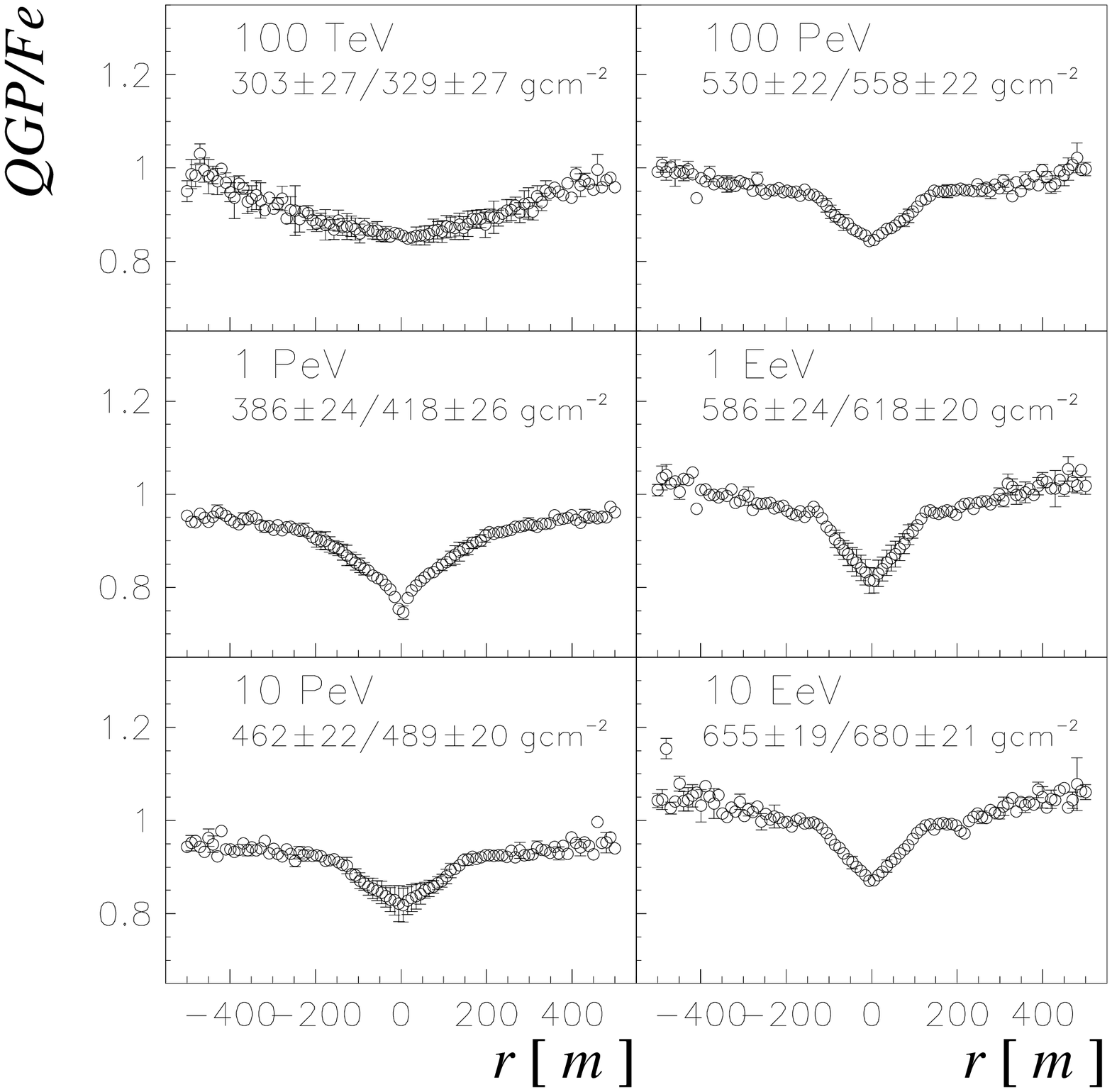}}
\vspace{-1.3cm}
\caption{Relative QGP/Fe ground level lateral distribution of Cherenkov 
photons.
Average values of QGP and Fe shower maxima are indicated.}
\label{F04}
\vspace{-0.8cm}
\end{figure}
In Fig.\ref{F03} the average ground level densities and ratios of total 
numbers of Cherenkov photons produced along the QGP and Fe showers 
are depicted as functions of the slant depth of their birth.

Average ratios of the ground level Cherenkov photon lateral 
densities from the QGP and Fe air cascades are depicted in 
Fig.\ref{F04}.  
The photon core densities of the QGP showers are typically by 
a factor of 1.2 smaller than the core densities of the Fe showers 
of the same primary energy. 
At very high primary energies, $E_{\scs{Fe}} \ge 1 \EeV$, 
the relative fraction of photons at outer parts of the QGP 
shower increases. 
Thus the QGP effects could be visible comparing the number 
of Cherenkov photons reaching the detector level relatively 
far from the shower core with all detected photons.

Any influence of the transverse momenta of the direct hadrons in 
the first interaction becomes negligible as in the case of very 
energetic primary the lateral profile of shower is dominated by 
interactions of particles at lower energies in the later part 
of the shower development.
The QGP showers behave similarly as ordinary hadronic showers 
initiated by the iron primary of somewhat lower energy.
However, the lateral spread of Cherenkov photons caused by the 
equilibrated QGP products is larger.
All these effects depend on both the primary energy and the distance 
between detector and shower maximum; the latter observable is also 
energy dependent.

\section{Conclusions}
\label{CO}

The lateral ground level Cherenkov photon densities of the extended 
air showers can be used to identify showers that follow 
the QGP transformation in the first interaction. 
It is worth noting that similar QGP related effects can be observed in lateral
distributions of high momenta muons detected underground~\cite{Rid1}.

\vspace{0.2cm}

This work was supported by the Ministry of Education of the Czech
Republic under contracts Nos. LN134 and LN00A006.



\end{document}